\pdfoutput=1

\documentclass[11pt]{article}

\usepackage[preprint]{acl}

\usepackage{amsmath}
\usepackage{times}
\usepackage{latexsym}
\usepackage{hyperref}
\usepackage{url}
\usepackage{booktabs}
\usepackage{multirow}
\usepackage{tcolorbox}
\usepackage{wrapfig}

\usepackage[T1]{fontenc}

\usepackage[utf8]{inputenc}

\usepackage{microtype}

\usepackage{inconsolata}

\usepackage{graphicx}

\title{T2A-Feedback: Improving Basic Capabilities of Text-to-Audio \\ Generation via Fine-grained AI Feedback}

\author{
   \vspace{-0.8cm} \\
    Zehan Wang$^{1}$\thanks{Equal Contribution.}, Ke Lei$^{1*}$, Chen Zhu$^{1}$, Jiawei Huang$^{1}$, Sashuai Zhou$^{1}$, \\ Luping Liu$^{2}$, Xize Cheng$^{1}$, Shengpeng Ji$^{1}$, Zhenhui Ye$^{1}$, Tao Jin$^{1}$, Zhou Zhao$^{1}$\thanks{Corresponding author.} \\ \\
   \vspace{-0.8cm} \\
   {$^1$}Zhejiang University; {$^2$}The University of Hong Kong \\
   \vspace{0.3cm}
}

\let\oldtwocolumn\twocolumn
\renewcommand\twocolumn[1][]{%
    \oldtwocolumn[{#1}{
    \begin{center}
    \vspace{-3.5\baselineskip}
    \includegraphics[width=0.9\linewidth]{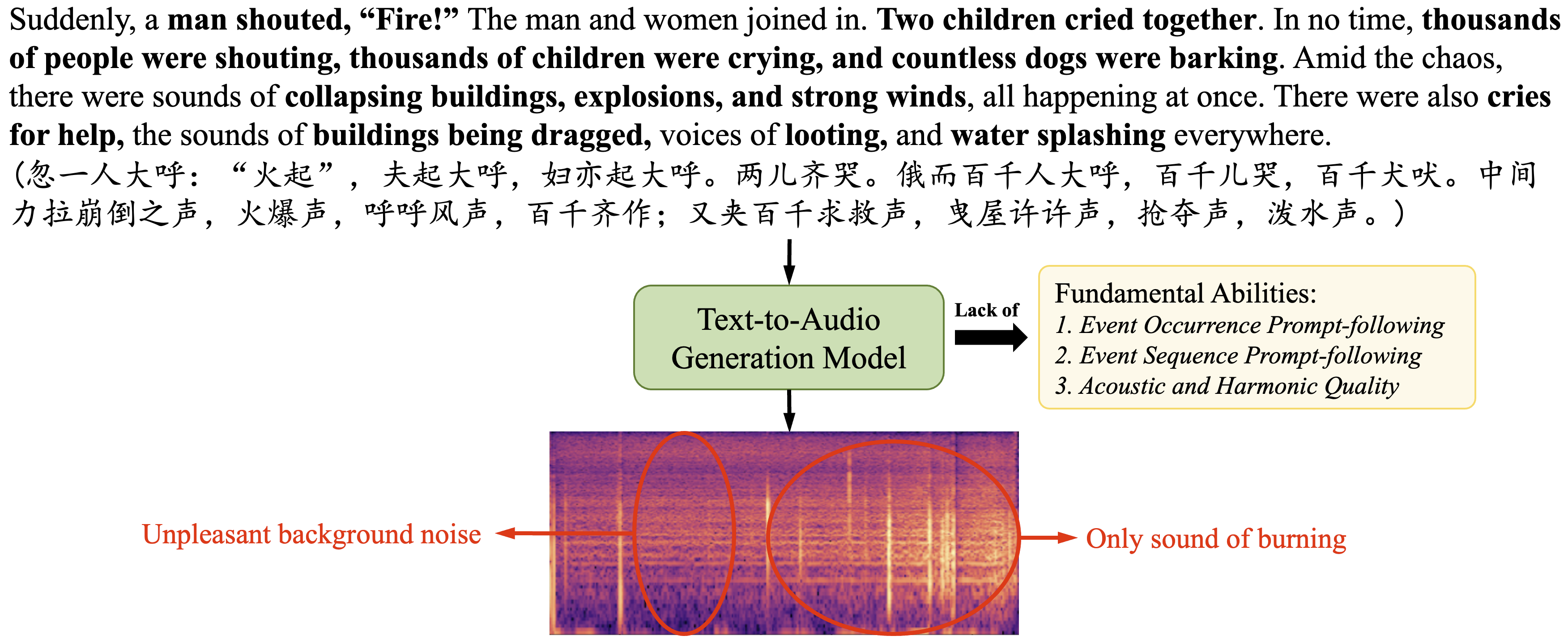}
    \captionof{figure}{The audio description is from a classic Chinese essay ``Kou Ji", which vividly depicts a performer using only vocal mimicry to recreate an entire dramatic scene. The existing Text-to-Audio generation model struggles to generate such narrative and multi-event audios. The generated audio often fails to contain all events in the described sequence while maintaining acoustic quality and harmony.}
    \label{fig:kouji}
        \end{center}
    }]
}

\begin{document}
\maketitle

\begin{abstract}
Text-to-audio (T2A) generation has achieved remarkable progress in generating a variety of audio outputs from language prompts. However, current state-of-the-art T2A models still struggle to satisfy human preferences for prompt-following and acoustic quality when generating complex multi-event audio. To improve the performance of the model in these high-level applications, we propose to enhance the basic capabilities of the model with AI feedback learning. First, we introduce fine-grained AI audio scoring pipelines to: 1) verify whether each event in the text prompt is present in the audio (\textbf{Event Occurrence Score}), 2) detect deviations in event sequences from the language description (\textbf{Event Sequence Score}), and 3) assess the overall acoustic and harmonic quality of the generated audio (\textbf{Acoustic\&Harmonic Quality}). We evaluate these three automatic scoring pipelines and find that they correlate significantly better with human preferences than other evaluation metrics. This highlights their value as both feedback signals and evaluation metrics. Utilizing our robust scoring pipelines, we construct a large audio preference dataset, \textbf{T2A-FeedBack}, which contains 41k prompts and 249k audios, each accompanied by detailed scores. Moreover, we introduce \textbf{T2A-EpicBench}, a benchmark that focuses on long captions, multi-events, and story-telling scenarios, aiming to evaluate the advanced capabilities of T2A models. Finally, we demonstrate how T2A-FeedBack can enhance current state-of-the-art audio model. With simple preference tuning, the audio generation model exhibits significant improvements in both simple (AudioCaps test set) and complex (T2A-EpicBench) scenarios. The project page is available at \url{https://T2Afeedback.github.io}

\end{abstract}

\section{Introduction}
Recent Text-to-Audio (T2A) generation models~\citep{huang2023make, huang2023make2, liu2023audioldm, liu2024audioldm, ghosal2023tango, majumder2024tango2, vyas2023audiobox} have made drastic performance improvements. By trained on massive audio-text data~\citep{gemmeke2017audio, fonseca2021fsd50k, chen2020vggsound, kim2019audiocaps}, these generative models learn to generate diverse sounds with a given language prompt. For audio generation, generating harmonious multi-event audio or describing a story with audio has important applications in music~\citep{agostinelli2023musiclm}, advertising, video-audio generation~\citep{luo2024diff, wang2024frieren}, etc.
However, as shown in Figure.~\ref{fig:kouji}, existing audio generation models are struggling to generate harmonious and high-quality audio from narrative and complex descriptions, which limits the potential for high-level applications.



The failure of the generated results is often demonstrated in three aspects: 1) cannot fully include all the events described, 2) cannot accurately follow the order of all the events described, and 3) cannot organize all the events harmoniously. Therefore, the model performance in multi-event scenarios is determined by its capabilities in these three fundamental aspects.



To improve the model's performance across more advanced applications, we focus on strengthening the audio generation model’s fundamental abilities. Inspired by feedback learning in large language models~\citep{ouyang2022training, bai2022training, touvron2023llama}, we propose creating an audio preference dataset centered on three abilities necessary for generating harmonic and complex audio: 1) \textbf{Event Occurrence Prompt-Following}, 2) \textbf{Event Sequence Prompt-Following}, and 3) \textbf{Acoustic\&Harmonic Quality}. Based on the preference information, we can refine the model’s core abilities, resulting in better results in both simple and challenging scenarios.

However, due to the scarcity of audio data and the challenges of annotating the scale of user preferences, it is difficult to collect massive audio preference datasets that only rely on human annotators. To fill this void, we explore using AI feedback~\citep{cui2023ultrafeedback, lee2023rlaif,  yuan2024self, burns2023weak} in text-to-audio generation, utilizing AI models to rank audios instead of relying on human annotators. Compared to manual annotation, automating the data collection and annotation process reduces the cost of obtaining audio preference data and enhances scalability.

Specifically, we develop three AI scoring pipelines to evaluate the generated audio in a fine-grained and holistic manner, corresponding to three core capabilities:
\begin{itemize}
    \item \textit{Event Occurrence Score}: To specifically check whether each event occurs in, we calculate the audio-text semantic matching score for each described event separately. A lower score suggests that the corresponding event might be absent from the audio.
    \item \textit{Event Sequence Score}: To verify the correctness of event order, we analyze the start and end times of each event and compare them with the event order outlined in the text prompt. A higher score implies a greater similarity between the event sequences in caption and audio.
    \item \textit{Acoustics\&Harmonic Quality}: 
    Drawing inspiration from aesthetic scoring methods used in image quality scoring, we manually annotate acoustic and harmonic quality for audio samples. These data are then used to train an automatic acoustic\&harmonic predictor.
\end{itemize}

We confirm that our three scoring functions show a stronger correlation with human evaluations compared to existing automatic audio evaluation methods~\citep{wu2023large, xie2024picoaudio}. Consequently, in addition to their application in ranking preference data, these scoring functions can be used as evaluation metrics that more effectively capture human preferences across different aspects.

Leveraging these advanced AI scoring pipelines, we establish a comprehensive data collection and annotation framework. As a result, we construct \textbf{T2A-Feedback}, a large audio preference dataset comprising 41,627 captions and 249,762 generated audios, each annotated with detailed scores.

Furthermore, to evaluate the higher-level capabilities of text-to-audio models in multi-event scenarios, we introduce a more challenging benchmark, \textbf{T2A-EpicBench}, which features longer, more imaginative, and story-telling captions for audio generation. We enhance the advanced text-to-audio diffusion model, Make-an-Audio 2~\citep{huang2023make2}, with T2A-Feedback. Our results show that using T2A-Feedback not only effectively improves the basic capabilities of the model in simple AudioCaps benchmark, but also emergently improves the performance in complex T2A-EpicBench.

\section{Related Work}
\subsection{Text-to-Audio Generation}
Text-to-audio generation is an emerging field that aims to convert textual descriptions into corresponding audio outputs. Existing text-to-audio generation methods can be divided into two categories: Diffusion-based and Language model-based. Diffusion-based techniques have gained prominence for generating high-quality, realistic audio by modeling the process of denoising. These methods, like Make-an-Audio~\citep{huang2023make, huang2023make2}, AudioLDM~\citep{liu2023audioldm, liu2024audioldm}, Tango~\citep{ghosal2023tango, majumder2024tango2}, start with random noise and iteratively refine it to produce coherent audio over a series of denoising steps. On the other hand, Language model-based methods~\citep{borsos2023audiolm, agostinelli2023musiclm, cideron2024musicrl} tokenize audios as acoustic discrete tokens, and predict the tokens within an auto-regressive model conditioned on text inputs.

The above models acquire the ability to generate diverse audio by training on large-scale audio-text datasets. However, current datasets like AudioSet~\citep{gemmeke2017audio}, AudioCaps~\citep{kim2019audiocaps}, and FSD50k~\citep{fonseca2021fsd50k} only provide tag-level annotations or short captions. As a result, when processing long, detailed language prompts, existing models often produce low-quality, noisy outputs and struggle to accurately follow the text. Due to the difficulty of annotating detailed audio captions, scaling rich and accurate audio descriptions remains a challenge. In this work, we focus on enhancing the model’s basic abilities in event occurrence, sequence, and harmony, thereby improving its performance in both simple scenarios and advanced applications.

\subsection{Perference Tuning with Human\&AI Feedback}
Tuning generative models according to human preferences has emerged as a standard practice for improving the quality of outputs. By tuning with feedback information on different aspects, the model can be improved and aligned with human preferences in corresponding aspects. Traditionally, this preference data used for tuning relied heavily on human evaluators who rank multiple generated results, assessing their quality based on various criteria such as relevance, coherence, and aesthetic value~\citep{bai2022training, touvron2023llama, ouyang2022training, kirstain2023pick, liang2024rich, wu2023human, cideron2024musicrl}. 

While effective, manual human annotation is costly and time-consuming, which greatly hampers the scalability of preference tuning across more diverse generative tasks. To address the difficulty, more recent developments have focused on leveraging pre-trained AI models to automate the process of scoring generated content~\citep{cui2023ultrafeedback, lee2023rlaif, yuan2024self, burns2023weak}. Such an AI feedback approach has achieved impressive improvements in large language models.

Recently, some studies have attempted preference fine-tuning in text-to-audio generation models. One recent paper related to our work, Tango2~\citep{majumder2024tango2}, utilizes contrastive language-audio pre-training (CLAP)~\citep{wu2023large} to rank audio generated by the Tango model. However, CLAP can only evaluate the global alignment between audio and text but falls short in assessing the fine-grained details, like detailed event occurrence, sequence, and harmony. In this paper, we construct more robust AI audio scoring pipelines with fine-grained recognition ability. Our method shows a much stronger correlation with human preference and the constructed dataset brings significant improvement to the current text-to-audio generation model.

\subsection{Text-to-Audio Evaluation Metric}
Existing evaluation metrics for audio generation, such as FAD and IS, assess audio distributions but cannot evaluate the quality of individual samples. Additionally, many studies rely on similarity scores from the CLAP model to assess global audio-text semantic alignment. PicoAudio~\citep{xie2024picoaudio} uses a text-to-audio grounding model~\citep{xu2024towards} to detect audio segments based on language prompts. However, there remains a lack of fine-grained evaluation methods for assessing detailed event occurrence, sequencing, and acoustic quality. Our research fills this gap by creating robust audio AI scoring pipelines, that show a strong correlation with humans, and significantly surpass alternative methods.

\section{T2A-Feedback}
\label{sec:method}
In this section, we first dive into the three AI audio scoring pipelines: (i) Event Occurrence Prompt-following, in Section.~\ref{sec:EOS}; (ii) Event Sequence Prompt-following, in Section.~\ref{sec:ESS}; (iii) Acoustic Quality, in Section.~\ref{sec:AHQ}. We then describe the specific data generation and sorting method for the T2A-Feedback dataset in Section.~\ref{sec:data}.

\begin{figure}
    \centering
    \includegraphics[width=1.04\linewidth]{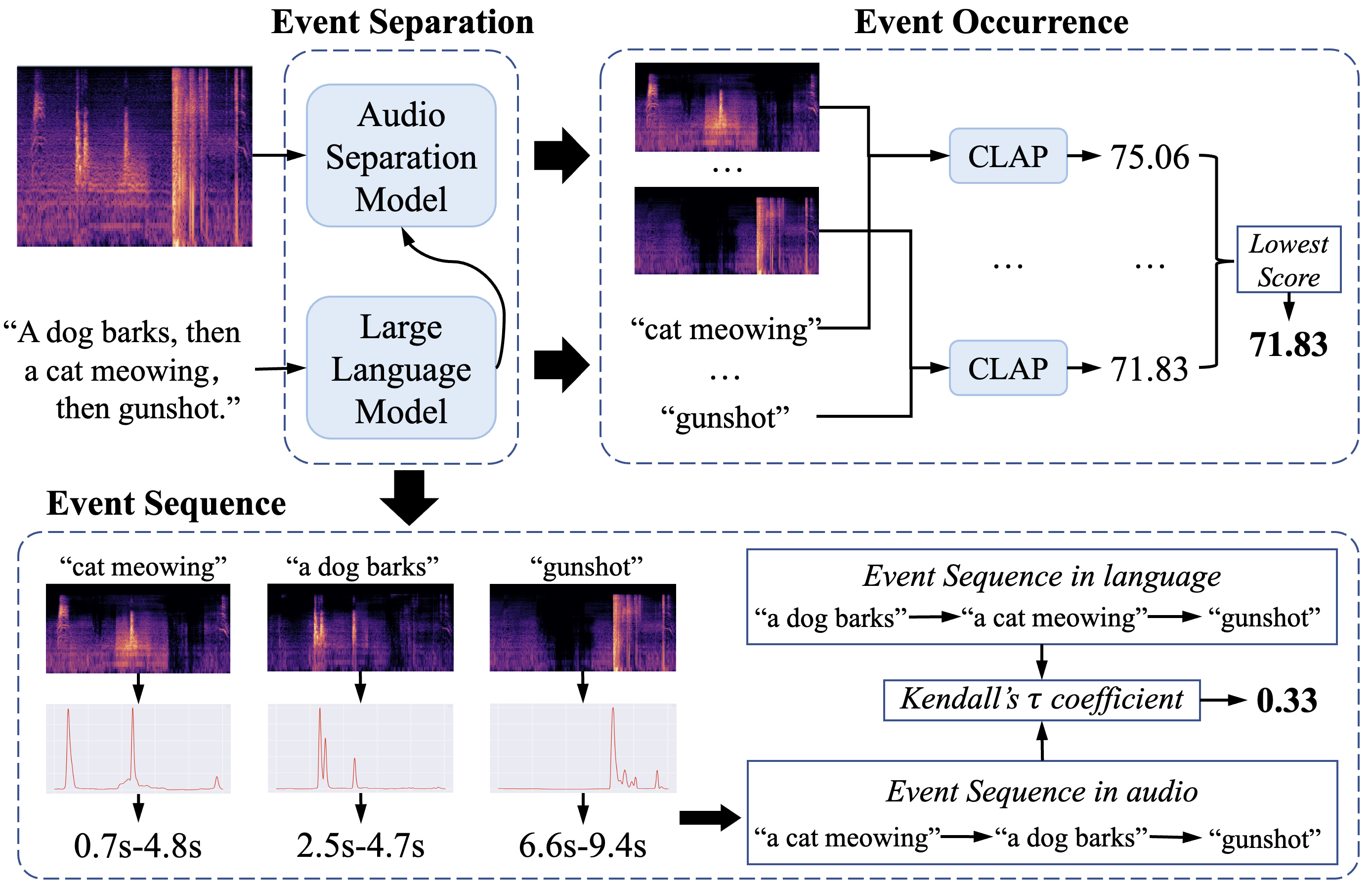}
    \caption{The overview of event occurrence and sequence scoring pipelines.}
    \label{fig:model}
\end{figure}

\subsection{Events Occurrence Prompt-following}
\label{sec:EOS}
Generating audio that accurately reflects the events described in a given prompt is the fundamental requirement of prompt-following. However, when multiple events are included in the text description, current text-to-audio generation models often struggle to generate each event precisely. To improve the generation model's event occurrence prompt-following ability, we first build an AI pipeline to determine the occurrence of events in audio.


Previous methods primarily utilize contrastive language-audio pre-training (CLAP)~\citep{wu2023large} over the audios and language descriptions to assess their semantic relevance. However, in multi-event scenarios, the sentence-level matching score struggles to identify event-level misalignment, and can not pinpoint which specific events are present and which are not, as shown in Figure.~\ref{fig:score_comparison}. To accurately identify misaligned events, we propose to measure the audio-text semantic alignment at the event-level. To this end, we first separate the language description and audio into basic events, as shown in the ``Event Separation" part of Figure.~\ref{fig:model}. Specifically, we utilize a large language model (LLM)~\citep{jiang2023mistral} to decompose descriptions into event captions according to the described order. Meanwhile, we employ an advanced audio separation model~\citep{liu2023separate} to segment the audio into event-level sub-audios based on these event captions. By calculating the similarity between each event-level description and its corresponding sub-audio in CLAP space, we can gain clearer insights into the specific aligned and misaligned events.

To encourage the models to comprehensively generate all described events, for each audio-text pair, we select the lowest value among all event-level audio-text matching scores as the \textbf{Event Occurrence Score}. For audios generated from the same caption, a higher score indicates that the audio is more likely to contain all the described events.

\subsection{Events Sequence Prompt-following}
\label{sec:ESS}
In addition to generating all events, whether these events occur in the temporal order described in the caption is also a crucial aspect of prompt-following. Some recent work attempts to detect the sequence of events in audio. Tango2~\citep{majumder2024tango2} computes the CLAP matching score between the temporal description and corresponding audios, but we find the sentence-level CLAP score is not sensitive to the temporal description in captions, as demonstrated in Figure.~\ref{fig:score_comparison} and Table.~\ref{tab:ess}. On the other hand, PicoAudio~\citep{xie2024picoaudio} employs audio grounding model~\citep{xu2024towards} to detect audio segments. However, due to the limitation of the training scale, the generalization performance of the audio grounding model is limited.

To robustly analyze audio event sequences, we propose a new pipeline for event sequence analysis. Similar to event occurrence, we first use the LLM and audio separation model to extract event-level descriptions and their corresponding sub-audios. For each separated audio track, we determine the event's start and end times based on volume levels. Specifically, we normalize the volume to a range of [0,1], and the period where the normalized volume exceeds a certain threshold is identified as the event's duration. 

In multi-event scenarios, there are multiple complex temporal relationships. To comprehensively assess the temporal alignment between the language prompt and the generated audio, and to specifically identify which temporal relationships are accurate and which are misaligned, we employ Kendall's $\tau$ coefficient. This widely used non-parametric statistic measures rank correlation between two variables. Considering $n$ events and their $n(n-1)$ event pairs, we use LLM to analyze the relationships between each event pair in the language description and extract the event sequence in the audio based on the starting time of each event. The \textbf{Events Sequence Score} (e.g., Kendall's $\tau$ coefficient between event sequences in language and audio) is calculated as:
\begin{equation}
    \tau = \dfrac{C-D}{n(n-1)}
\end{equation}
where $C$ represents the number of concordant event pairs between the description and the audio, $D$ denotes the number of discordant ones. Higher $\tau$ indicates a greater alignment of the event sequence in the generated audio with the text description. Specifically, $\tau=1$ signifies that the event sequence in the generated audio is identical to the language description, while $\tau=-1$ indicates that the sequences are completely reversed.

\subsection{Acoustic\&Harmonic Quality}
\label{sec:AHQ}
In addition to generating all events accurately following the language prompt, organically integrating different events to create a pleasant-sounding effect is also one of the basic capabilities. However, current audio generation models often produce low-quality and noisy results. 

To alleviate this challenge, we first develop an audio acoustic\&harmonic quality predictor.
Inspired by the image aesthetic predictor in \cite{schuhmann2022laion}, we first manually score audio samples on a scale from 1 to 4 according to their quality. Three annotators independently score the audio samples according to the same criteria, and samples with consistent scores are accepted as training data. Detailed scoring criteria is provided in Appendix.

Using the human-annotated data, we train a linear predictor on the top of CLAP audio embeddings. With the high-quality pre-trained representation, we find that, akin to aesthetic score predictors for images, a small amount of annotated data can yield a generalized subjective quality predictor. Specifically, we train the acoustic predictor with 2,000 meticulously annotated audio samples using cross-entropy loss. The output of the predictor is termed the \textbf{Acoustic\&Harmonic Quality}.

\subsection{Preference Data Generation}
\label{sec:data}
To generate diverse and comprehensive audio, we first augment the text prompts used for audio generation. 
We begin with the captions from the training set of the large-scale audio-text dataset, AudioCaps. By employing an LLM, we decompose these captions into fundamental event descriptions and calculate their semantic similarity within the CLAP space to filter out non-overlapping, basic event descriptions. Then, we prompt the LLM with randomly selected events to create varied and natural multi-event audio descriptions, with explicit temporal ordering. Finally, we combine the enhanced 3,769 captions with the 37,858 captions from the training set of AudioCaps, serving as the prompt source for audio generation.

As highlighted in~\cite{cui2023ultrafeedback}, diversity is crucial for preference datasets. To mitigate the potential bias of using a single audio generation model and to enhance the generalization of the generated data, we employ three advanced audio generation models: Make-an-Audio2, AudioLDM2, and Tango2. Each model generates 2 audio per caption, resulting in a total of 6 audio files for each caption. In summary, we produce 249,762 audios from 41,627 descriptions. For audios generated from the same captions, we combine three rankings of each score to derive the overall ranking.

The histogram plots of the scores on all the generated audios are shown in Appendix. The distribution of Event Occurrence Scores and Acoustic\&Harmonic Quality is similar to a Gaussian distribution. Since most descriptions contain one or two sequential events, Event Sequence Scores are concentrated between -1 and 1. As noted in~\cite{liang2024rich}, this discriminative score distribution ensures a balanced ratio of negative to positive samples, enabling effective preference tuning.




\section{T2A-EpicBench}
Current text-to-audio generation models are mainly evaluated and compared on the AudioCaps test set. However, the captions in AudioCaps are generally short and simple, averaging 10.3 words per sentence. Specifically, 17\% of the captions feature only a single event, and 44\% contains two events. This is not enough to assess the model’s capabilities in more advanced applications involving detailed, multi-event, and narrative-style audio generation.

To fill this gap, we propose \textbf{T2A-EpicBench}, consisting of 100 detailed, multi-event, and story-telling captions. Each caption averages 54.8 words and 4.2 events, with 86\% containing four events and the remainder featuring five or more.  Initially, we manually write 10 detailed captions, then used them as in-context examples to prompt LLM for generating the remaining captions. All 100 captions are manually reviewed for accuracy. Several examples from T2A-EpicBench are included in the Appendix.

\section{Experiment}
\label{sec:experiment}
\subsection{Analysis of AI Scoring Pipelines}


\subsubsection{Quantitative analysis}
\label{sec:analysis}
\paragraph{Evaluation of Event Occurrence Score (EOS)} 
To evaluate the scoring model's capability in recognizing whether audios contain all the events described in the text, we propose a missing event recognition task. We construct distracting captions by adding random event descriptions to the ground-truth captions. This task challenges models to distinguish the ground-truth caption from the constructed interference captions. The test sets of AudioCaps (3,701 samples), Clotho (5,225 samples), and MusicCaps (4,434 samples) are employed for evaluation.

We mainly compare our EOS with CLAP and PANNs. The caption with the higher matching score to the audio is considered as the prediction. For the audio tagging model, PANNs, we match the top 5 recognized audio categories with open-domain descriptions. As shown in Table~\ref{tab:eos}, our EOS score showcases a notable advantage over baselines on all the benchmarks, demonstrating the superiority of event-level audio-text matching in identifying whether all events are correctly contained in audios.


\begin{table}[t]
\resizebox{\linewidth}{!}{
\begin{tabular}{cccc}
\toprule
 & \textbf{AudioCaps} & \textbf{Clotho} & \textbf{MusicCaps} \\ \midrule
Random Guess & 50.0\% & 50.0\% & 50.0\% \\
CLAP & 77.5\% & 86.4\% & 69.4\%   \\
PANNs & 82.0\% & 79.9\% & 56.1\% \\ \midrule
EOS(ours) & \textbf{90.9\%} &  \textbf{90.4\%} & \textbf{99.8\%}  \\ \bottomrule
\end{tabular}}
\vspace{-0.5\baselineskip}
\caption{Results of event occurrence recognition}
\label{tab:eos}
\end{table}

\begin{table}[t]
\resizebox{\linewidth}{!}{
\begin{tabular}{cccccc}
\toprule
& \multicolumn{2}{c}{\textbf{Two Events}} & \multicolumn{2}{c}{\textbf{Three Events}} & \multirow{2}{*}{Correlation} \\
\cmidrule(lr){2-3} \cmidrule(lr){4-5}  & Acc & F1 & Acc & F1 & \\ \midrule
CLAP & 49.6\%  & - & 53.7\% & - & - \\
PicoAudio & 71.6\%&	0.787 & 51.3 & 0.574 & 0.30 \\ \midrule
ESS$_{0.1}$ & \textbf{79.6\%}	&0.814	& 54.2 & \textbf{0.606} & 0.43 \\ 
ESS$_{0.3}$ & 79.1\%&	\textbf{0.851}	& \textbf{57.6} & 0.587 & \textbf{0.52} \\
ESS$_{0.5}$ & 78.0\%&	0.769	& 56.7 & 0.535 & 0.52 \\ \bottomrule
\end{tabular}}
\vspace{-0.5\baselineskip}
\caption{Results of event sequence recognition. ESS$_{0.x}$ stands for using $0.x$ as volume thresholds.}
\label{tab:ess}
\end{table}

\begin{figure}[t]
    \centering
    \includegraphics[width=1\linewidth]{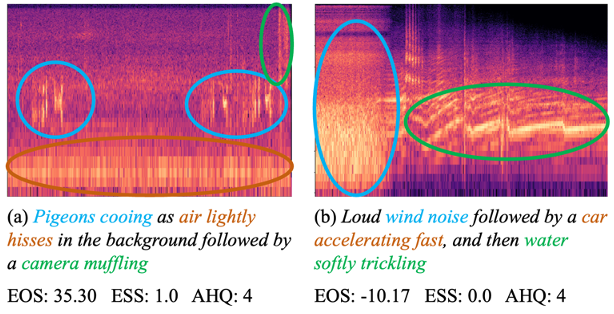}
    \vspace{-1.5\baselineskip}
    \caption{Visualization of the predicted scores from our AI scoring pipeline. We highlight the first, second, and third events described in the captions using blue, brown, and green, respectively.}
    \label{fig:score_example}
\end{figure}

\begin{figure}[t]
    \centering
    \includegraphics[width=1\linewidth]{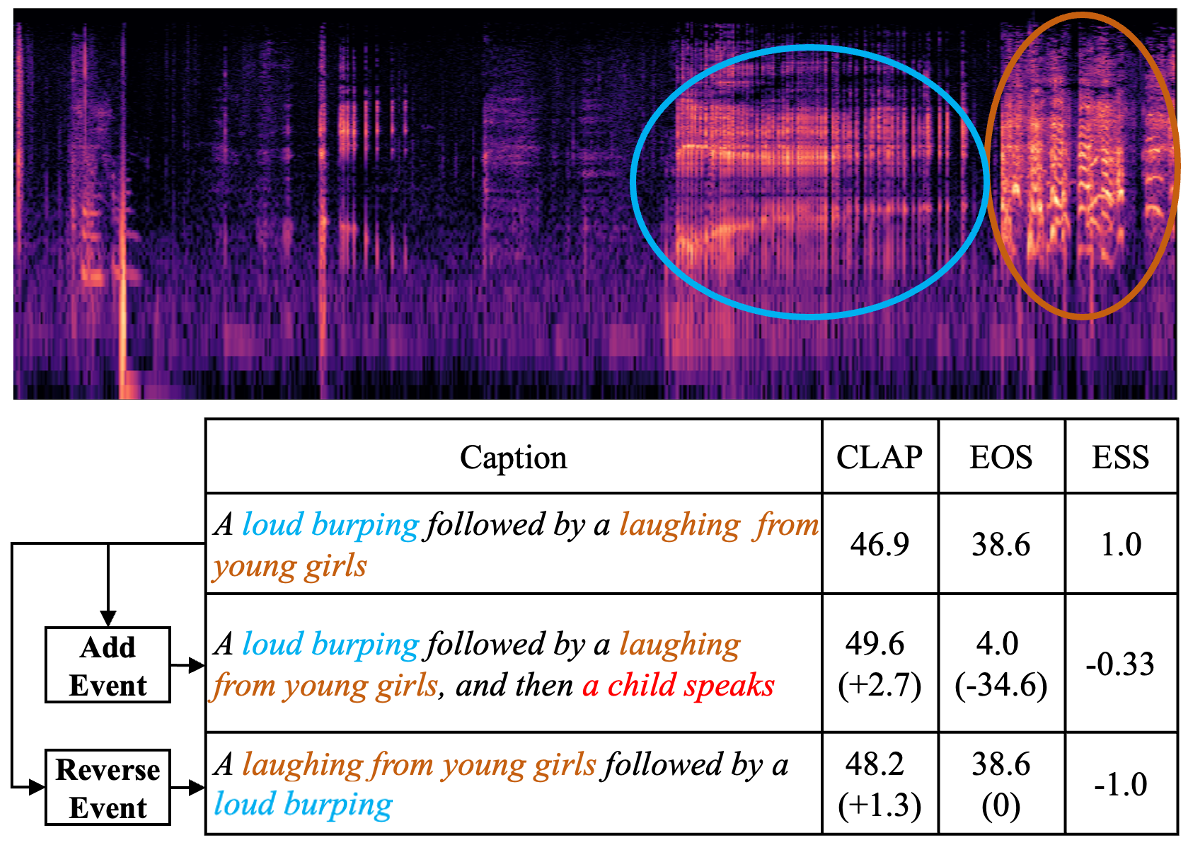}
    \vspace{-1.5\baselineskip}
    \caption{Qualitative comparison between CLAP scores and EOS/ESS scores reveals distinct sensitivities to misalignment. By adding or reversing events in the ground-truth caption, the captions become misaligned with the audio in terms of event occurrence and sequence.}
    
    \label{fig:score_comparison}
\end{figure}

\paragraph{Evaluation of Event Sequence Score (ESS)} 
To verify the ability to distinguish the alignment of event sequences in text and audio, we collect 450 two-event and 200 three-event samples from PicoAudio's data, and reverse the events orders in the description as interference caption. Using this dataset, we compare different methods by calculating the accuracy of recognizing the ground-truth description versus the interference description, and by evaluating the Segment F1 Score~\citep{mesaros2016metrics} for detecting the start and end times of each audio event. Moreover, we manually annotate temporal order alignment for 100 audios generated from our temporal-enhanced captions and compute the correlation between different methods and humans.

The results of event sequences are provided in Table.~\ref{tab:ess}. We compare ESS with CLAP score and the audio grounding model~\citep{xu2024towards} used by PicoAudio~\citep{xie2024picoaudio}. Compared to baselines, our method distinguishes the ground-truth caption from the distracting one more accurately and achieves higher F1 scores in detecting the start and end times of events in audio. More importantly, our method shows a much stronger correlation to human annotations. 

Additionally, we investigate various volume thresholds used to determine the duration of each event. In Table~\ref{tab:ess}, we test thresholds of 0.1, 0.3, and 0.5. ESS consistently outperforms other methods across most settings, with 0.3 providing the optimal results and thus chosen as the default setting.


\paragraph{Evaluation of Acoustic\&Harmonic Quality (AHQ)}
To validate our acoustic\&harmonic predictor, we independently annotate 100 additional audios as a test set. The correlation between the model predictions and human labels on the test set is 0.786, showing strong generalization ability and high consistency with human preferences.


Moreover, we explore building the Acoustic\&Harmonic Predictor on top of various pre-trained audio models and evaluate how well each variant correlates with human preferences. The correlation of the predictor built on CLAP~\citep{wu2023large} (0.79) outperforms those based on self-supervised models like AudioMAE~\citep{huang2022masked} (0.61) and BEAT~\citep{chen2022beats} (0.52). Similarly, the image aesthetics predictor~\citep{schuhmann2022laion} is built on the CLIP model~\citep{ilharco_gabriel_2021_5143773}. This advantage may stem from the alignment with language, resulting in better semantic discrimination.


\subsubsection{Qualitative Analysis}
We show some example predictions from our scoring pipelines in Figure.~\ref{fig:score_example}, where our methods can specifically identify the misaligned event, the out-of-order event order, and the disharmony between events in the audio. Moreover, we provide the qualitative comparison between our EOS and ESS with the single CLAP score, in Figure.~\ref{fig:score_comparison}. For the ground-truth audio-caption pairs from AudioCaps, we perturb the captions by adding an event or shuffling the order of events. We find that the CLAP score is not sensitive to these perturbations and even yields a higher score with the incorrect, perturbed caption. In contrast, our EOS and ESS scores more accurately reflect the alignment between audio and text regarding event occurrence and event order.

\begin{table}[t]
\centering
\setlength\tabcolsep{3pt}
\resizebox{\linewidth}{!}{
\begin{tabular}{ccccccccc}
\toprule

\multicolumn{2}{c}{} & FAD$\downarrow$ & KL$\downarrow$ & IS$\uparrow$ & CLAP$\uparrow$ & EOS.$\uparrow$ & ESS.$\uparrow$ & AQ.$\uparrow$ \\ \midrule
\multicolumn{2}{c}{Make an Audio 2} & \textbf{1.82} & 1.44 & 10.03 & 69.97 & 42.05 &	0.53  &  2.33 \\ \midrule
\multicolumn{9}{c}{Preference Tuning} \\ \midrule
\multirow{2}{*}{Audio-Alpaca} & RAFT & {1.93} &	\textbf{1.29}	& 10.37 & 72.23 & 44.85 & 0.53 & 2.45 \\
 & DPO & 3.20 &	{1.24}	& \textbf{12.27} & 72.36 & 44.42 & 0.55 & 2.14 \\ \midrule
\multirow{2}{*}{\begin{tabular}[c]{@{}c@{}}T2A-Feedback\\ (ours)\end{tabular}} & RAFT & 2.29 & 1.33 & {11.66} & 73.10 & 45.53 & 0.51 & 2.50 \\
 & DPO & 2.64 &	{1.31} &	11.35 & \textbf{74.00} & \textbf{49.58} & \textbf{0.57} & \textbf{2.57} \\ \midrule
\end{tabular}}
\vspace{-0.5\baselineskip}
\caption{Evaluation results on AudioCaps. The EOS, ESS and AHQ represent the Event Occurrence Score, Event Sequence Score and Acoustic\&Harmonic Quality, respectively.}
\vspace{-0.5\baselineskip}
\label{tab:audiocaps}
\end{table}

\begin{table}[t]
\centering
\setlength\tabcolsep{3pt}
\resizebox{\linewidth}{!}{
\begin{tabular}{cccccccc}
\toprule
\multicolumn{2}{l}{} & \multicolumn{3}{c}{AI Scoring} & \multicolumn{3}{c}{Human Scoring} \\
\cmidrule(lr){3-5} \cmidrule(lr){6-8}
\multicolumn{2}{c}{} & win$_{EOS}$ & win$_{ESS}$ & win$_{AHQ}$ & win$_{EOS}$ & win$_{ESS}$ & win$_{AHQ}$ \\ \midrule
\multicolumn{2}{c}{Make an Audio 2} & -  (14.21) & -  (0.03) & -  (1.96) & - &	- & - \\ \midrule
\multicolumn{8}{c}{Preference Tuning} \\ \midrule
\multirow{2}{*}{Audio-Alpaca} & RAFT & 53\%(15.73) & 51\%(0.04) & 42\%(1.69) & 57\% & 54\% & 53\% \\
 & DPO & 55\%(16.87) & 52\%(0.03) & 49\%(1.96) & 65\% & \textbf{64\%}  & 59\% \\\midrule
\multirow{2}{*}{\begin{tabular}[c]{@{}c@{}}T2A-Feedback\\ (ours)\end{tabular}} & RAFT & 52\%(15.85) & 52\%(0.05)	& \textbf{54\%(2.14)}  & 61\%  & 57\%  & 61\% \\
 & DPO & \textbf{58\%(19.96)} & \textbf{64\%(0.13)} & 52\%(2.10) & \textbf{68\%}  & {62\%} & \textbf{68\%}  \\ \midrule
\end{tabular}}
\vspace{-0.5\baselineskip}
\caption{Evaluation results on T2A-EpicBench. The win$_{EOS}$, win$_{ESS}$ and win$_{AHQ}$ represent the win rates of tuned models over the original model in terms of Event Occurrence, Event Sequence and Acoustic\&Harmonic Quality, respectively.}
\vspace{-0.5\baselineskip}
\label{tab:epicbench}
\end{table}

\subsection{Analysis of Preference Tuning}
To demonstrate the effect of T2A-Feedback dataset in improving audio generation model, we finetuning the advanced text-to-audio model, Make-an-Audio 2~\citep{huang2023make2}, with two preference training methods: Direct Preference Optimization (DPO)~\citep{wallace2024diffusion} and Reward rAnked FineTuning (RAFT)~\citep{dong2023raft}. Another audio preference dataset, Audio-Alpace, proposed by~\cite{majumder2024tango2} is the main baseline for comparison. Both the widely-used AudioCaps and the new T2A-EpicBench are used as benchmarks, corresponding to applications in simple and complex scenarios respectively.

\begin{figure*}[t]
    \centering
    \vspace{-1\baselineskip}
    \textit{a). AudioCaps} \\
    \includegraphics[width=0.98\linewidth]{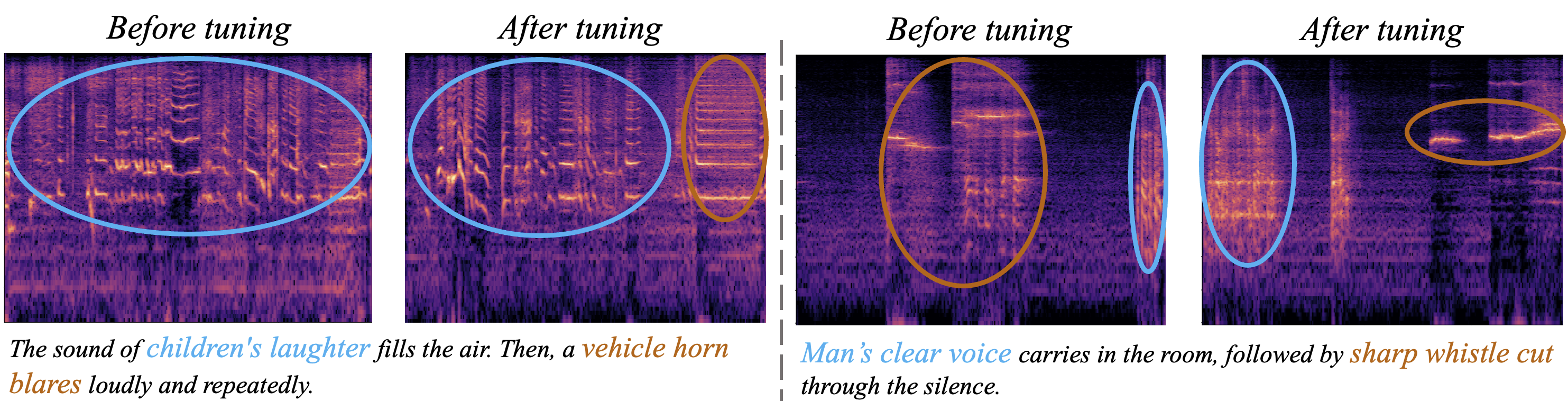}  \\
    \textit{b). T2A-EpicBench}  \\
    \includegraphics[width=0.98\linewidth]{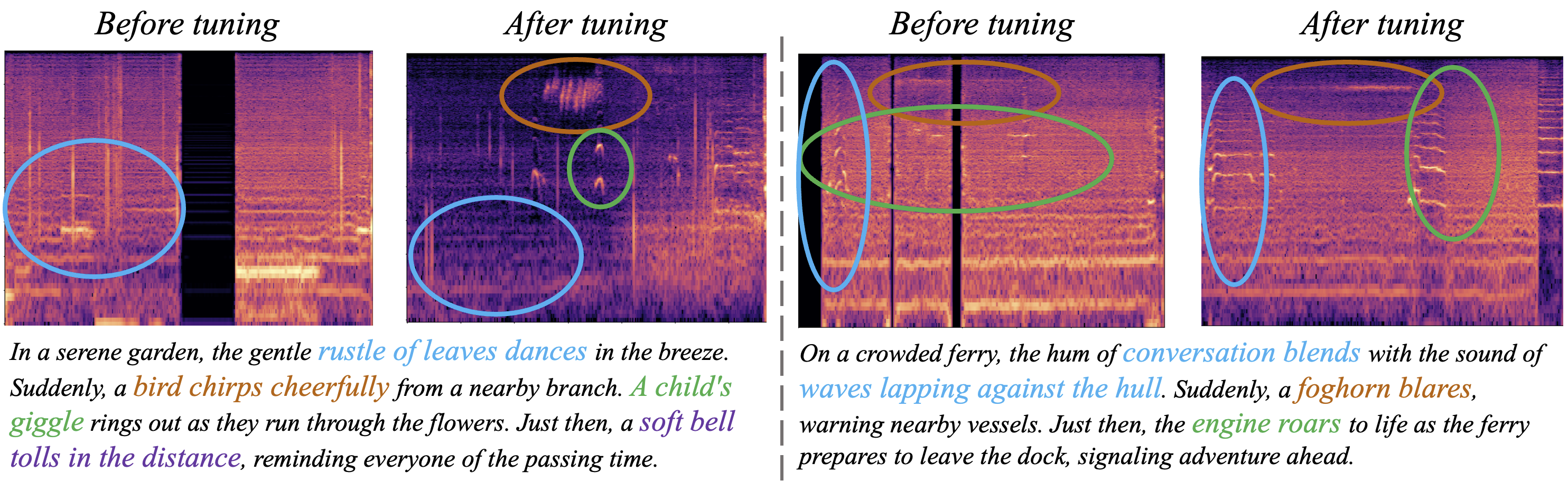}
    \vspace{-0.5\baselineskip}
    \caption{Visualization of the impact of preference tuning with T2A-Feedback.}
    \vspace{-1\baselineskip}
    \label{fig:vis}
\end{figure*}

\subsubsection{Quantitative Results on AudioCaps}
The classical automated metrics (FAD, KL, IS, and CLAP), as well as our three new scores (EOS, ESS, and AHQ) are employed to quantitatively evaluate and compare different model variants. 

The quantitative results are provided in Table.~\ref{tab:audiocaps}. FAD, KL, and IS assess audio fidelity by evaluating the distribution of the generated audio. For these metrics, both the preference dataset and training methods result in similar overall improvements. CLAP is commonly used to measure the semantic alignment between the input prompt and the generated audio. While both Audio-Alpaca and T2A-Feedback improve the CLAP score, T2A-Feedback yields greater gains. 

Moreover, as analyzed in Section.~\ref{sec:analysis}, the proposed EOS and ESS are more accurate than CLAP in judging event occurrence and event sequence, and AHQ shows a strong correlation to human preference in acoustic and harmony. We calculate the three scores for different model variants to evaluate audio generation results more accurately and comprehensively. The significantly better results across these three metrics demonstrate that T2A-Feedback yields far greater improvements compared to Audio-Alpaca, and the DPO method outperforms RAFT in our setting.

\subsubsection{Quantitative Results on T2A-EpicBench}
Since there are no ground-truth audios for the long and story-telling text prompts in T2A-EpicBench, we primarily measure the win rate of preference-tuned models against the original model outputs across three key areas: event occurrence, event sequence, and acoustic \& harmonic quality. In addition to scoring the generated audio with our AI pipeline, we conduct a user study where two human annotators evaluate and select the better output based on each criterion.

The results on T2A-EpicBench, are illustrated in Table.~\ref{tab:epicbench}, indicate that Audio-Alpaca provides only marginal improvements in handling detailed captions and multi-event scenarios, whereas T2A-Feedback significantly and comprehensively enhances the model's performance. 

It is worth noting that T2A-Feedback does not include long audio descriptions. The average word count per caption in T2A-Feedback is 9.6, which is considerably shorter than the 54.8 average word number of T2A-EpicBench prompts, and even shorter than Audio-Alpaca’s 10.2 words per caption. T2A-Feedback does not directly provide additional long caption data, and the 65\% average win rate in the user study reinforces that by focusing on improving the basic capabilities, the audio generation model can emergently learn to handle more complex long-text and multi-event scenarios.

\subsubsection{Qualitative Findings}

To better demonstrate the effectiveness of preference tuning on T2A-Feedback, we visualize some examples of tuning the original model on our T2A-Feedback with the DPO method in Figure.~\ref{fig:vis}. For the examples of short captions in Figure.~\ref{fig:vis}a, while both models before and after fine-tuning can produce clean audio, the fine-tuned model successfully generates all events in the described order. In the more challenging case from T2A-EpicBench, the original model often generates noisy, low-quality audio, making it difficult to distinguish the events. Preference tuning on T2A-Feedback, as shown in Figure.~\ref{fig:vis}b, significantly reduces background noise and generates audio that more faithfully captures both events and their orders.




\section{Conclusion}
In this paper, we build AI scoring pipelines to evaluate three fundamental capabilities of audio generation: Event Occurrence Prompt-following, Event Sequence Prompt-following, and Acoustics\&Harmonic Quality. Using these automatic evaluation metrics, which are highly correlated with human preferences, we build a large-scale audio preference dataset, \textbf{T2A-Feedback}. Experimentally, the three scores demonstrate a strong correlation to human preferences, which highlights its potential to better evaluate text-to-audio generation models. To assess the model's ability in complex multi-event scenarios, we propose a new challenging benchmark, \textbf{T2A-EpicBench}, which requires models to generate detailed and narrative audios. Using our T2A-Feedback to tune the audio generation model effectively improves its capabilities in the three basic  capabilities and achieves better performance in both simple (AudioCaps) and complex (T2A-EpicBench) scenarios.

\section*{Limitation}
Automatically generating high-quality and harmonious audio from detailed, narrative, and multi-event scenarios remains a long-term goal. The performance of the audio generation model depends on both the pre-training phase and the post-training phase (fine-tuning and feedback learning). To fully address the challenge of generating coherent audio for long narrative prompts, improvements are needed across the entire process.

\bibliography{custom}

\begin{thebibliography}{39}
\providecommand{\natexlab}[1]{#1}

\bibitem[{Agostinelli et~al.(2023)Agostinelli, Denk, Borsos, Engel, Verzetti, Caillon, Huang, Jansen, Roberts, Tagliasacchi et~al.}]{agostinelli2023musiclm}
Andrea Agostinelli, Timo~I Denk, Zal{\'a}n Borsos, Jesse Engel, Mauro Verzetti, Antoine Caillon, Qingqing Huang, Aren Jansen, Adam Roberts, Marco Tagliasacchi, et~al. 2023.
\newblock Musiclm: Generating music from text.
\newblock \emph{arXiv preprint arXiv:2301.11325}.

\bibitem[{Bai et~al.(2022)Bai, Jones, Ndousse, Askell, Chen, DasSarma, Drain, Fort, Ganguli, Henighan et~al.}]{bai2022training}
Yuntao Bai, Andy Jones, Kamal Ndousse, Amanda Askell, Anna Chen, Nova DasSarma, Dawn Drain, Stanislav Fort, Deep Ganguli, Tom Henighan, et~al. 2022.
\newblock Training a helpful and harmless assistant with reinforcement learning from human feedback.
\newblock \emph{arXiv preprint arXiv:2204.05862}.

\bibitem[{Borsos et~al.(2023)Borsos, Marinier, Vincent, Kharitonov, Pietquin, Sharifi, Roblek, Teboul, Grangier, Tagliasacchi et~al.}]{borsos2023audiolm}
Zal{\'a}n Borsos, Rapha{\"e}l Marinier, Damien Vincent, Eugene Kharitonov, Olivier Pietquin, Matt Sharifi, Dominik Roblek, Olivier Teboul, David Grangier, Marco Tagliasacchi, et~al. 2023.
\newblock Audiolm: a language modeling approach to audio generation.
\newblock \emph{IEEE/ACM transactions on audio, speech, and language processing}, 31:2523--2533.

\bibitem[{Burns et~al.(2023)Burns, Izmailov, Kirchner, Baker, Gao, Aschenbrenner, Chen, Ecoffet, Joglekar, Leike et~al.}]{burns2023weak}
Collin Burns, Pavel Izmailov, Jan~Hendrik Kirchner, Bowen Baker, Leo Gao, Leopold Aschenbrenner, Yining Chen, Adrien Ecoffet, Manas Joglekar, Jan Leike, et~al. 2023.
\newblock Weak-to-strong generalization: Eliciting strong capabilities with weak supervision.
\newblock \emph{arXiv preprint arXiv:2312.09390}.

\bibitem[{Chen et~al.(2020)Chen, Xie, Vedaldi, and Zisserman}]{chen2020vggsound}
Honglie Chen, Weidi Xie, Andrea Vedaldi, and Andrew Zisserman. 2020.
\newblock Vggsound: A large-scale audio-visual dataset.
\newblock In \emph{ICASSP 2020-2020 IEEE International Conference on Acoustics, Speech and Signal Processing (ICASSP)}, pages 721--725. IEEE.

\bibitem[{Chen et~al.(2022)Chen, Wu, Wang, Liu, Tompkins, Chen, and Wei}]{chen2022beats}
Sanyuan Chen, Yu~Wu, Chengyi Wang, Shujie Liu, Daniel Tompkins, Zhuo Chen, and Furu Wei. 2022.
\newblock Beats: Audio pre-training with acoustic tokenizers.
\newblock \emph{arXiv preprint arXiv:2212.09058}.

\bibitem[{Cideron et~al.(2024)Cideron, Girgin, Verzetti, Vincent, Kastelic, Borsos, McWilliams, Ungureanu, Bachem, Pietquin et~al.}]{cideron2024musicrl}
Geoffrey Cideron, Sertan Girgin, Mauro Verzetti, Damien Vincent, Matej Kastelic, Zal{\'a}n Borsos, Brian McWilliams, Victor Ungureanu, Olivier Bachem, Olivier Pietquin, et~al. 2024.
\newblock Musicrl: Aligning music generation to human preferences.
\newblock \emph{arXiv preprint arXiv:2402.04229}.

\bibitem[{Cui et~al.(2023)Cui, Yuan, Ding, Yao, Zhu, Ni, Xie, Liu, and Sun}]{cui2023ultrafeedback}
Ganqu Cui, Lifan Yuan, Ning Ding, Guanming Yao, Wei Zhu, Yuan Ni, Guotong Xie, Zhiyuan Liu, and Maosong Sun. 2023.
\newblock Ultrafeedback: Boosting language models with high-quality feedback.
\newblock \emph{arXiv preprint arXiv:2310.01377}.

\bibitem[{Dong et~al.(2023)Dong, Xiong, Goyal, Zhang, Chow, Pan, Diao, Zhang, Shum, and Zhang}]{dong2023raft}
Hanze Dong, Wei Xiong, Deepanshu Goyal, Yihan Zhang, Winnie Chow, Rui Pan, Shizhe Diao, Jipeng Zhang, Kashun Shum, and Tong Zhang. 2023.
\newblock Raft: Reward ranked finetuning for generative foundation model alignment.
\newblock \emph{arXiv preprint arXiv:2304.06767}.

\bibitem[{Fonseca et~al.(2021)Fonseca, Favory, Pons, Font, and Serra}]{fonseca2021fsd50k}
Eduardo Fonseca, Xavier Favory, Jordi Pons, Frederic Font, and Xavier Serra. 2021.
\newblock Fsd50k: an open dataset of human-labeled sound events.
\newblock \emph{IEEE/ACM Transactions on Audio, Speech, and Language Processing}, 30:829--852.

\bibitem[{Gemmeke et~al.(2017)Gemmeke, Ellis, Freedman, Jansen, Lawrence, Moore, Plakal, and Ritter}]{gemmeke2017audio}
Jort~F Gemmeke, Daniel~PW Ellis, Dylan Freedman, Aren Jansen, Wade Lawrence, R~Channing Moore, Manoj Plakal, and Marvin Ritter. 2017.
\newblock Audio set: An ontology and human-labeled dataset for audio events.
\newblock In \emph{2017 IEEE international conference on acoustics, speech and signal processing (ICASSP)}, pages 776--780. IEEE.

\bibitem[{Ghosal et~al.(2023)Ghosal, Majumder, Mehrish, and Poria}]{ghosal2023tango}
Deepanway Ghosal, Navonil Majumder, Ambuj Mehrish, and Soujanya Poria. 2023.
\newblock Text-to-audio generation using instruction-tuned llm and latent diffusion model.
\newblock \emph{arXiv preprint arXiv:2304.13731}.

\bibitem[{Huang et~al.(2023{\natexlab{a}})Huang, Ren, Huang, Yang, Ye, Zhang, Liu, Yin, Ma, and Zhao}]{huang2023make2}
Jiawei Huang, Yi~Ren, Rongjie Huang, Dongchao Yang, Zhenhui Ye, Chen Zhang, Jinglin Liu, Xiang Yin, Zejun Ma, and Zhou Zhao. 2023{\natexlab{a}}.
\newblock Make-an-audio 2: Temporal-enhanced text-to-audio generation.
\newblock \emph{arXiv preprint arXiv:2305.18474}.

\bibitem[{Huang et~al.(2022)Huang, Xu, Li, Baevski, Auli, Galuba, Metze, and Feichtenhofer}]{huang2022masked}
Po-Yao Huang, Hu~Xu, Juncheng Li, Alexei Baevski, Michael Auli, Wojciech Galuba, Florian Metze, and Christoph Feichtenhofer. 2022.
\newblock Masked autoencoders that listen.
\newblock \emph{Advances in Neural Information Processing Systems}, 35:28708--28720.

\bibitem[{Huang et~al.(2023{\natexlab{b}})Huang, Huang, Yang, Ren, Liu, Li, Ye, Liu, Yin, and Zhao}]{huang2023make}
Rongjie Huang, Jiawei Huang, Dongchao Yang, Yi~Ren, Luping Liu, Mingze Li, Zhenhui Ye, Jinglin Liu, Xiang Yin, and Zhou Zhao. 2023{\natexlab{b}}.
\newblock Make-an-audio: Text-to-audio generation with prompt-enhanced diffusion models.
\newblock In \emph{International Conference on Machine Learning}, pages 13916--13932. PMLR.

\bibitem[{Ilharco et~al.(2021)Ilharco, Wortsman, Wightman, Gordon, Carlini, Taori, Dave, Shankar, Namkoong, Miller, Hajishirzi, Farhadi, and Schmidt}]{ilharco_gabriel_2021_5143773}
Gabriel Ilharco, Mitchell Wortsman, Ross Wightman, Cade Gordon, Nicholas Carlini, Rohan Taori, Achal Dave, Vaishaal Shankar, Hongseok Namkoong, John Miller, Hannaneh Hajishirzi, Ali Farhadi, and Ludwig Schmidt. 2021.
\newblock \href {https://doi.org/10.5281/zenodo.5143773} {Openclip}.
\newblock If you use this software, please cite it as below.

\bibitem[{Jiang et~al.(2023)Jiang, Sablayrolles, Mensch, Bamford, Chaplot, Casas, Bressand, Lengyel, Lample, Saulnier et~al.}]{jiang2023mistral}
Albert~Q Jiang, Alexandre Sablayrolles, Arthur Mensch, Chris Bamford, Devendra~Singh Chaplot, Diego de~las Casas, Florian Bressand, Gianna Lengyel, Guillaume Lample, Lucile Saulnier, et~al. 2023.
\newblock Mistral 7b.
\newblock \emph{arXiv preprint arXiv:2310.06825}.

\bibitem[{Kim et~al.(2019)Kim, Kim, Lee, and Kim}]{kim2019audiocaps}
Chris~Dongjoo Kim, Byeongchang Kim, Hyunmin Lee, and Gunhee Kim. 2019.
\newblock Audiocaps: Generating captions for audios in the wild.
\newblock In \emph{Proceedings of the 2019 Conference of the North American Chapter of the Association for Computational Linguistics: Human Language Technologies, Volume 1 (Long and Short Papers)}, pages 119--132.

\bibitem[{Kirstain et~al.(2023)Kirstain, Polyak, Singer, Matiana, Penna, and Levy}]{kirstain2023pick}
Yuval Kirstain, Adam Polyak, Uriel Singer, Shahbuland Matiana, Joe Penna, and Omer Levy. 2023.
\newblock Pick-a-pic: An open dataset of user preferences for text-to-image generation.
\newblock \emph{Advances in Neural Information Processing Systems}, 36:36652--36663.

\bibitem[{Lee et~al.(2023)Lee, Phatale, Mansoor, Mesnard, Ferret, Lu, Bishop, Hall, Carbune, Rastogi et~al.}]{lee2023rlaif}
Harrison Lee, Samrat Phatale, Hassan Mansoor, Thomas Mesnard, Johan Ferret, Kellie Lu, Colton Bishop, Ethan Hall, Victor Carbune, Abhinav Rastogi, et~al. 2023.
\newblock Rlaif: Scaling reinforcement learning from human feedback with ai feedback.
\newblock \emph{arXiv preprint arXiv:2309.00267}.

\bibitem[{Liang et~al.(2024)Liang, He, Li, Li, Klimovskiy, Carolan, Sun, Pont-Tuset, Young, Yang et~al.}]{liang2024rich}
Youwei Liang, Junfeng He, Gang Li, Peizhao Li, Arseniy Klimovskiy, Nicholas Carolan, Jiao Sun, Jordi Pont-Tuset, Sarah Young, Feng Yang, et~al. 2024.
\newblock Rich human feedback for text-to-image generation.
\newblock In \emph{Proceedings of the IEEE/CVF Conference on Computer Vision and Pattern Recognition}, pages 19401--19411.

\bibitem[{Liu et~al.(2023{\natexlab{a}})Liu, Chen, Yuan, Mei, Liu, Mandic, Wang, and Plumbley}]{liu2023audioldm}
Haohe Liu, Zehua Chen, Yi~Yuan, Xinhao Mei, Xubo Liu, Danilo Mandic, Wenwu Wang, and Mark~D Plumbley. 2023{\natexlab{a}}.
\newblock Audioldm: Text-to-audio generation with latent diffusion models.
\newblock \emph{arXiv preprint arXiv:2301.12503}.

\bibitem[{Liu et~al.(2024)Liu, Yuan, Liu, Mei, Kong, Tian, Wang, Wang, Wang, and Plumbley}]{liu2024audioldm}
Haohe Liu, Yi~Yuan, Xubo Liu, Xinhao Mei, Qiuqiang Kong, Qiao Tian, Yuping Wang, Wenwu Wang, Yuxuan Wang, and Mark~D Plumbley. 2024.
\newblock Audioldm 2: Learning holistic audio generation with self-supervised pretraining.
\newblock \emph{IEEE/ACM Transactions on Audio, Speech, and Language Processing}.

\bibitem[{Liu et~al.(2023{\natexlab{b}})Liu, Kong, Zhao, Liu, Yuan, Liu, Xia, Wang, Plumbley, and Wang}]{liu2023separate}
Xubo Liu, Qiuqiang Kong, Yan Zhao, Haohe Liu, Yi~Yuan, Yuzhuo Liu, Rui Xia, Yuxuan Wang, Mark~D Plumbley, and Wenwu Wang. 2023{\natexlab{b}}.
\newblock Separate anything you describe.
\newblock \emph{arXiv preprint arXiv:2308.05037}.

\bibitem[{Luo et~al.(2024)Luo, Yan, Hu, and Zhao}]{luo2024diff}
Simian Luo, Chuanhao Yan, Chenxu Hu, and Hang Zhao. 2024.
\newblock Diff-foley: Synchronized video-to-audio synthesis with latent diffusion models.
\newblock \emph{Advances in Neural Information Processing Systems}, 36.

\bibitem[{Majumder et~al.(2024)Majumder, Hung, Ghosal, Hsu, Mihalcea, and Poria}]{majumder2024tango2}
Navonil Majumder, Chia-Yu Hung, Deepanway Ghosal, Wei-Ning Hsu, Rada Mihalcea, and Soujanya Poria. 2024.
\newblock Tango 2: Aligning diffusion-based text-to-audio generations through direct preference optimization.
\newblock \emph{arXiv preprint arXiv:2404.09956}.

\bibitem[{Mesaros et~al.(2016)Mesaros, Heittola, and Virtanen}]{mesaros2016metrics}
Annamaria Mesaros, Toni Heittola, and Tuomas Virtanen. 2016.
\newblock Metrics for polyphonic sound event detection.
\newblock \emph{Applied Sciences}, 6(6):162.

\bibitem[{Ouyang et~al.(2022)Ouyang, Wu, Jiang, Almeida, Wainwright, Mishkin, Zhang, Agarwal, Slama, Ray et~al.}]{ouyang2022training}
Long Ouyang, Jeffrey Wu, Xu~Jiang, Diogo Almeida, Carroll Wainwright, Pamela Mishkin, Chong Zhang, Sandhini Agarwal, Katarina Slama, Alex Ray, et~al. 2022.
\newblock Training language models to follow instructions with human feedback.
\newblock \emph{Advances in neural information processing systems}, 35:27730--27744.

\bibitem[{Schuhmann et~al.(2022)Schuhmann, Beaumont, Vencu, Gordon, Wightman, Cherti, Coombes, Katta, Mullis, Wortsman et~al.}]{schuhmann2022laion}
Christoph Schuhmann, Romain Beaumont, Richard Vencu, Cade Gordon, Ross Wightman, Mehdi Cherti, Theo Coombes, Aarush Katta, Clayton Mullis, Mitchell Wortsman, et~al. 2022.
\newblock Laion-5b: An open large-scale dataset for training next generation image-text models.
\newblock \emph{Advances in Neural Information Processing Systems}, 35:25278--25294.

\bibitem[{Touvron et~al.(2023)Touvron, Martin, Stone, Albert, Almahairi, Babaei, Bashlykov, Batra, Bhargava, Bhosale et~al.}]{touvron2023llama}
Hugo Touvron, Louis Martin, Kevin Stone, Peter Albert, Amjad Almahairi, Yasmine Babaei, Nikolay Bashlykov, Soumya Batra, Prajjwal Bhargava, Shruti Bhosale, et~al. 2023.
\newblock Llama 2: Open foundation and fine-tuned chat models.
\newblock \emph{arXiv preprint arXiv:2307.09288}.

\bibitem[{Vyas et~al.(2023)Vyas, Shi, Le, Tjandra, Wu, Guo, Zhang, Zhang, Adkins, Ngan et~al.}]{vyas2023audiobox}
Apoorv Vyas, Bowen Shi, Matthew Le, Andros Tjandra, Yi-Chiao Wu, Baishan Guo, Jiemin Zhang, Xinyue Zhang, Robert Adkins, William Ngan, et~al. 2023.
\newblock Audiobox: Unified audio generation with natural language prompts.
\newblock \emph{arXiv preprint arXiv:2312.15821}.

\bibitem[{Wallace et~al.(2024)Wallace, Dang, Rafailov, Zhou, Lou, Purushwalkam, Ermon, Xiong, Joty, and Naik}]{wallace2024diffusion}
Bram Wallace, Meihua Dang, Rafael Rafailov, Linqi Zhou, Aaron Lou, Senthil Purushwalkam, Stefano Ermon, Caiming Xiong, Shafiq Joty, and Nikhil Naik. 2024.
\newblock Diffusion model alignment using direct preference optimization.
\newblock In \emph{Proceedings of the IEEE/CVF Conference on Computer Vision and Pattern Recognition}, pages 8228--8238.

\bibitem[{Wang et~al.(2024)Wang, Guo, Huang, Huang, Wang, You, Li, and Zhao}]{wang2024frieren}
Yongqi Wang, Wenxiang Guo, Rongjie Huang, Jiawei Huang, Zehan Wang, Fuming You, Ruiqi Li, and Zhou Zhao. 2024.
\newblock Frieren: Efficient video-to-audio generation with rectified flow matching.
\newblock \emph{arXiv preprint arXiv:2406.00320}.

\bibitem[{Wu et~al.(2023{\natexlab{a}})Wu, Hao, Sun, Chen, Zhu, Zhao, and Li}]{wu2023human}
Xiaoshi Wu, Yiming Hao, Keqiang Sun, Yixiong Chen, Feng Zhu, Rui Zhao, and Hongsheng Li. 2023{\natexlab{a}}.
\newblock Human preference score v2: A solid benchmark for evaluating human preferences of text-to-image synthesis.
\newblock \emph{arXiv preprint arXiv:2306.09341}.

\bibitem[{Wu et~al.(2023{\natexlab{b}})Wu, Chen, Zhang, Hui, Berg-Kirkpatrick, and Dubnov}]{wu2023large}
Yusong Wu, Ke~Chen, Tianyu Zhang, Yuchen Hui, Taylor Berg-Kirkpatrick, and Shlomo Dubnov. 2023{\natexlab{b}}.
\newblock Large-scale contrastive language-audio pretraining with feature fusion and keyword-to-caption augmentation.
\newblock In \emph{ICASSP 2023-2023 IEEE International Conference on Acoustics, Speech and Signal Processing (ICASSP)}, pages 1--5. IEEE.

\bibitem[{Xie et~al.(2024)Xie, Xu, Wu, and Wu}]{xie2024picoaudio}
Zeyu Xie, Xuenan Xu, Zhizheng Wu, and Mengyue Wu. 2024.
\newblock Picoaudio: Enabling precise timestamp and frequency controllability of audio events in text-to-audio generation.
\newblock \emph{arXiv preprint arXiv:2407.02869}.

\bibitem[{Xu et~al.(2024)Xu, Ma, Wu, and Yu}]{xu2024towards}
Xuenan Xu, Ziyang Ma, Mengyue Wu, and Kai Yu. 2024.
\newblock Towards weakly supervised text-to-audio grounding.
\newblock \emph{arXiv preprint arXiv:2401.02584}.

\bibitem[{Yuan et~al.(2024)Yuan, Pang, Cho, Sukhbaatar, Xu, and Weston}]{yuan2024self}
Weizhe Yuan, Richard~Yuanzhe Pang, Kyunghyun Cho, Sainbayar Sukhbaatar, Jing Xu, and Jason Weston. 2024.
\newblock Self-rewarding language models.
\newblock \emph{arXiv preprint arXiv:2401.10020}.

\bibitem[{Zheng et~al.(2024)Zheng, Teng, Yang, Wang, Chen, Gu, Dong, Ding, and Tang}]{zheng2024cogview3}
Wendi Zheng, Jiayan Teng, Zhuoyi Yang, Weihan Wang, Jidong Chen, Xiaotao Gu, Yuxiao Dong, Ming Ding, and Jie Tang. 2024.
\newblock Cogview3: Finer and faster text-to-image generation via relay diffusion.
\newblock \emph{arXiv preprint arXiv:2403.05121}.

\end{thebibliography}

\newpage
\appendix

\section{Examples from T2A-EpicBench}
\begin{tcolorbox}[colback=gray!10,
                  colframe=black,
                  width=0.5\textwidth,
                  arc=1mm, auto outer arc,
                  boxrule=0.5pt,
                  box align=center,
                 ]
                 
1. At a lively beach, the waves crash rhythmically against the shore, providing a soothing melody. Suddenly, a seagull caws overhead, drawing attention from sunbathers. Children‘s giggles fill the air as they splash in the water. Just then, a distant drumbeat starts, adding a festive atmosphere to the scene.
\\
\\
2. In a vibrant classroom, the teacher's voice resonates as she explains a lesson. Suddenly, a pencil rolls off a desk and clatters to the floor, causing a brief distraction. A student whispers a joke, provoking a wave of giggles. Just then, the school bell rings, signaling the end of the period.
\\
\\
3. In a busy city street, the honking of cars creates a chaotic symphony. Suddenly, a bicycle bell rings sharply as a cyclist weaves through traffic. The murmur of pedestrians chatting fills the air, blending with the distant sound of street performers playing music. Just then, the sound of footsteps approaches, adding to the urban rhythm.
\\
\\
4. At a busy construction site, the sound of drills and saws fills the air, creating a symphony of labor. Suddenly, a heavy beam falls with a loud thud, causing workers to pause. A whistle blows, signaling a break, and conversations buzz among the crew. Just then, a truck backs up, beeping as it arrives.
\\
\\
5. In a vibrant downtown area, the honking of cars creates a chaotic symphony. Suddenly, a street vendor shouts out their specials, trying to attract customers. The laughter of people enjoying a nearby café adds warmth to the urban sounds. Just then, a bus rumbles past, its engine growling as it continues.
\\
\\
6. In a vibrant market, the chatter of vendors fills the air as they hawk their goods. Suddenly, a loud crash echoes as a stack of crates falls over, causing startled gasps. A nearby musician strums a guitar, trying to restore the upbeat mood. Just then, a child squeals with delight, tugging at their parent's hand to explore further.
\end{tcolorbox}

\section{Implementation Details}
\label{sec:details}
\paragraph{Audio Generation} During the audio generation process in T2A-Feedback, all models are set to 100 denoising steps with the DDIM scheduler, and classifier-free guidance is configured at 4.0.

\paragraph{Training Details} For Acoustic\&Harmonic Predictor, we train an extra two-layer MLP projector on the top of CLAP audio representations using Cross Entropy(CE) loss. The predictor is trained using the Adam optimizer with a learning rate of 1e-2.5 for 6 epochs on 1,000 manually annotated data. For preference tuning, we employ the AdamW optimizer with a learning rate of 1e-5 for both DPO and RAFT strategy, and train one epoch for both Audio-Alpaca and T2A-Feedback.

\section{Other models on T2A-EpicBench}
The performance of AudioLDM 2 and Tango 2 on T2A-EpicBench is as follows:

\begin{table}[t]
    \centering
    \begin{tabular}{cccc}
    \toprule
        & EOS & ESS & AHQ \\ \midrule
        Make an Audio 2 & 14.21	& 0.03	& 1.96 \\
        AudioLDM2 &16.35& 	0.04	& 1.98 \\
        Tango2 & 19.42& 	0.07& 	\textbf{2.11}\\ \midrule
        \begin{tabular}[c]{@{}c@{}}Make an Audio 2 + \\ T2A-Feedback (DPO)\end{tabular} & \textbf{19.96}	& \textbf{0.13}& 	2.10 \\ \bottomrule
    \end{tabular}
    \caption{Results of AudioLDM2 and Tango2 on our T2A-EpicBench.}
    \label{tab:more_model}
\end{table}

\begin{figure*}[t]
    \centering
    \includegraphics[width=1\linewidth]{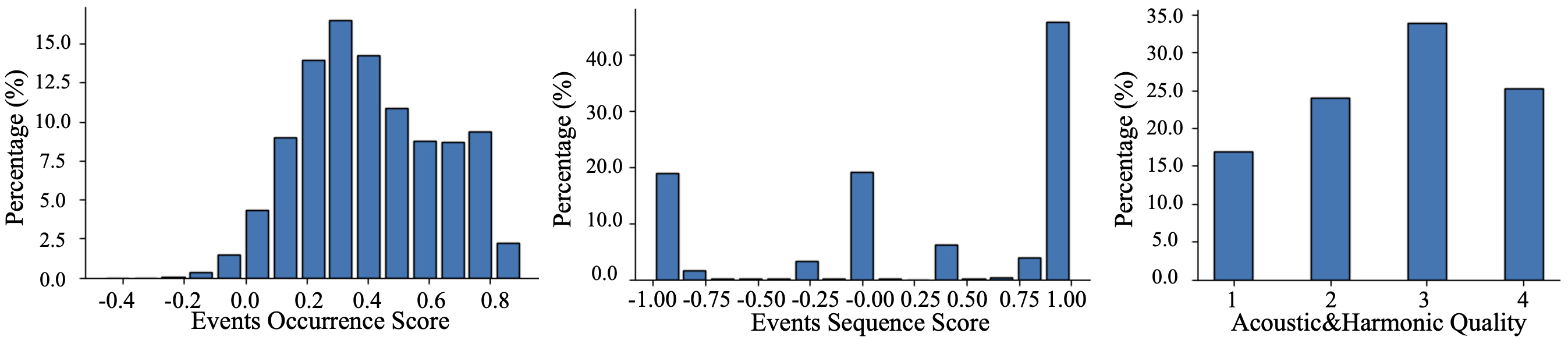}
    \caption{Histograms of three different scores in T2A-Feedback.}
    \label{fig:histograms}
\end{figure*}

As shown in Table.~\ref{tab:more_model}, the improvements observed across Make-an-audio 2, AudioLDM2, and Tango2 on EpicBench align with their inherent capabilities, with newer and more advanced models achieving better results. This indirectly validates the robustness and effectiveness of our benchmark and AI-based scoring pipeline.

Moreover, we observed that although the Make-an-audio 2 model does not perform well on EpicBench initially, it achieves the best performance after feedback alignment with T2A-Feedback. This highlights the practicality and significance of our dataset.

\section{Study about Lowest Score for EOS}
\begin{table}[]
    \centering
    \begin{tabular}{cccc}
    \toprule
         & \textbf{AudioCaps} & \textbf{Clotho} & \textbf{MusicCaps} \\ \midrule
      Average &  89.3	&88.8&	99.8 \\
      Lowest &   \textbf{90.9}&	\textbf{90.4}&	\textbf{99.8}\\ \bottomrule
    \end{tabular}
    \caption{Comparison between selecting lowest or average score for event occurrence score}
    \label{tab:lowest}
\end{table}
We tested the effect of selecting the average score and the lowest score among all matching scores for event occurrence judgment in Table.~\ref{tab:lowest}. We find that using the lowest score can better distinguish the caption with extra events for current audio-text datasets. According to the statistical results, we empirically select the lowest score for event occurrence.

\section{Negative Effect to FAD Score}
FAD and FID estimate the mean and covariance of two sample groups in a high-dimensional feature space and calculate their similarity. A negative correlation between FAD (FID) and subjective metrics is widely observed in the text-to-image and text-to-audio generations. The study Pick-a-Pic~\cite{kirstain2023pick} for text-to-image feedback learning has discussed this phenomenon, suggesting that it may be correlated to the classifier-free guidance scale mechanism. Larger classifier-free guidance scales tend to produce more vivid samples, which humans generally prefer, but deviate from the distribution of ground truth samples in the test set, resulting in worse (higher) FID (FAD) scores.

More specifically, this phenomenon is witnessed in Tables 1 and 2 of CogView3~\cite{zheng2024cogview3} (text-to-image method) and Table 3 of Tango2~\cite{majumder2024tango2} (text-to-audio method), where models achieve higher human preference scores but worse FID (FAD) scores. The negative correlation between FID (FAD) and subjective scores, as consistently shown by previous methods, appears to be an expected outcome when aligning generative models with human preferences.

\section{Statistic of Each Score}
We provide the histogram maps of three different scores in Figure.~\ref{fig:histograms}.

\section{Agreement between AHQ Annotations}
We provide the agreement between Acoustic\&Harmonic Quality (AHQ) annotations and predictions in Table.~\ref{tab:agreement}. All the annotators exhibit an agreement rate of over 70\% with the majority vote, which demonstrates the reliability of our annotation process.

\begin{table}[]
    \centering
    \resizebox{\linewidth}{!}{
    \begin{tabular}{c|cccc}
    \toprule
         & A-1 & A-2 & A-3 & Majority \\ \midrule
       Predictor  & 70.31\%&	68.75\%	&62.50\%	&73.44\% \\
       A-1  &	-	&64.06\%	&68.75\%	&74.33\% \\
       A-2  & 64.06\%&	-	&65.63\%&	71.88\% \\
       A-3  & 68.75\%	&65.63\% &	-	&70.31\% \\ \bottomrule
    \end{tabular}}
    \caption{Agreement between AHQ annotations and predictions on 100 testing samples. A-1, A-2 and A-3 are three human annotators. ``Majority" stands for the agreement between each judge and the other three judges’s majority votes.}
    \label{tab:agreement}
\end{table}

\section{Potential Risks}
Since our audio data is generated by the model based on the text, its content is mainly determined by the provided text. Therefore, if using generation models without safety checkers, offensive and unsafe content may be generated. In our work, we checked the content of the text prompt to ensure that the generated data does not contain offensive content.

\end{document}